\documentclass[aps,preprint,prb]{revtex4}
\usepackage{graphicx}
\usepackage{epsfig}
\usepackage{amsmath}
\usepackage{graphics}
\begin{document}

\author{M. F. Gelin}
\author{D. S. Kosov}

\affiliation{Department of Chemistry and Biochemistry,
 University of Maryland, 
 College Park, 
 MD 20742 }

\title{Angular momentum dependent friction slows down rotational relaxation
under non-equilibrium conditions}

\begin{abstract}
It has recently been shown that relaxation of 
the rotational energy of hot non-equlibrium photofragments (i) slows down significantly
with the increase of their initial rotational temperature  and (ii) differs dramatically from the relaxation of the equilibrium
rotational energy correlation function, manifesting thereby breakdown of the linear response description 
{[}Science 311, 1907 (2006){]}. We demonstrate that this phenomenon
may be caused by the angular momentum dependence of rotational friction.
We have developed the generalized Fokker-Planck
equation whose rotational friction depends upon angular momentum algebraically. 
The calculated rotational correlation functions
correspond well to their counterparts obtained via molecular dynamics simulations
in a broad range of initial non-equilibrium conditions. It
is suggested that the angular momentum dependence of friction should
be taken into account while describing rotational relaxation far from
equilibrium.
\end{abstract}
\maketitle

\section{Introduction}

For several decades, computer simulations, theoretical models,
and polarization experiments have been shaping and deepening our understanding
of how molecules reorient in liquids and solutions. Most 
of these studies deal with equilibrium molecular ensembles. Much
less is known about molecular reorientation under non-equilibrium
conditions. 

The study of photodissociation in a condensed phase offers such a
possibility. Indeed, the incipient photofragments are highly non-equilibrium,
and their rotational excitation is determined by the excited state potential energy surface  
of the parent molecule. Traditionally, polarization photodissociation
experiments (both in the steady state and in the time domain) have
been performed in the gas phase under collision-free conditions. \cite{zew94,zew01}
Recently, a number of
\char`\"{}real time\char`\"{} measurements \cite{hoc91,anf95,anf04,lim05,ham05,ruh93,wie93,wie96,zew96,hoc96,hoc97,hoc97a,voh98,voh99,voh99a,vol99,voh00,voh00a,nib01,bra03}
and computer simulations \cite{kar91,wil89,ben93,ben96,ger94} has been 
reported, where the anisotropy decay of diatomic photoproducts was
studied in the condensed phase. Both classical \cite{gel00,gel01}
and quantum \cite{gel02} models of the anisotropy decay in the dissipative
ensemble of photofragments have also been developed. 

So far, theoretical efforts have primarily been focused on studying
experimental observables, i.e. polarization time-dependent transients
(anisotropies). Very recently, an account of combined theoretical
and experimental study of  rotational and orientational relaxation
of linear CN photofragments under highly non-equilibrium conditions
has been published. \cite{str06,str06a} The authors report a number of extremely
interesting and sometimes unexpected phenomena, which are quite difficult
to comprehend within the existing theoretical paradigms. The authors demonstrate,   
in particularly, that relaxation of the rotational energy of hot non-equlibrium photofragments (i) slows down significantly
with the increase of their initial rotational temperature  and (ii) differs dramatically from the relaxation of the equilibrium
rotational energy correlation function, manifesting thereby breakdown of the linear response description.
What are the physical origins of the long 
rotational relaxation times and the linear response failure?  These are the questions which interest us here. 
We suggest that the angular momentum dependence
of the rotational friction is responsible for the observed behaviour.
We offer an explanation why this effect does not manifest itself 
under equilibrium conditions, while it becomes pivotal far from equilibrium. 

The structure of our paper is the following. The analysis of the problem
of rotational relaxation within the exact generalized master equations,
as well as the formal solution of these equations in terms of eigenfunctions
is presented in Sec. 2. The generalized Fokker-Planck equation (FPE)
with the angular momentum dependent friction is developed
in Sec. 3. In this section, the FPE with algebraic friction 
is solved for various rotational correlation functions (CFs), which
are compared with the CFs simulated in Ref. \cite{str06} Our main
findings are summarized in the Conclusion.

Note that the reduced variables are used throughout the article: time,
angular momentum and energy are measured in units of $\sqrt{I/(k_{B}T})$,
$\sqrt{Ik_{B}T}$ and $k_{B}T$, respectively. Here $k_{B}$ is the
Boltzmann constant, $T$ is the temperature of the equilibrium bath
molecules, and $I$ is the moment of inertia of the photofragment,
so that $\tau_{r}=\sqrt{I/(k_{B}T})$ is the averaged period of its
free rotation. For CN at $120$K, $\tau_{r}=$$0.3$ps.

\section{General equations }

Let us consider an ensemble of photofragments coupled to a heat bath.
We assume that the photofragments interact with the bath molecules,
but do not interact with each other. We suppose that at the time moment
$t=0$ the joint probability density in the photofragment + bath phase
space can be written as a product of the equilibrium distribution
for the bath molecules and a certain non-equilibrium distribution $\rho_{ne}(\mathbf{J})$
for the photofragments. \cite{foot1} Restricting our consideration to linear/spherical
photofragments, we can apply the projection operator technique to
the many body Liouville equation and derive the exact rotational master
equation for photofragments \cite{fre75,eva78,gel98}

\begin{equation}
\partial_{t}\rho(\mathbf{J},\mathbf{\Omega},t)=-i\hat{\Lambda}\rho(\mathbf{J},\mathbf{\Omega},t)-\int_{0}^{t}dt'\hat{C}(t-t')\rho(\mathbf{J},\mathbf{\Omega},t').\label{kin1}\end{equation}
 Here $\rho(\mathbf{J},\mathbf{\Omega},t)$ is the probability density
function, $\mathbf{J}$ is the angular momentum of the photofragment
in its molecular frame, $\mathbf{\Omega}$ are the Euler angles which
specify orientation of the molecular frame with respect to the laboratory
one. The free-rotor Liouville operator describes the angular momentum
driven reorientation,\begin{equation}
\hat{\Lambda}=\mathbf{J}\hat{\mathbf{L}},\label{str1}\end{equation}
 $\hat{\mathbf{L}}$ being the angular momentum operator in the molecular
frame. The relaxation operator obeys normalization 

\begin{equation}
\int d\mathbf{J}d\mathbf{\Omega}\hat{C}(t)=0,\label{norm}\end{equation}
which insures the conservation of probability ($\int d\mathbf{J}d\mathbf{\Omega}\rho(\mathbf{J},\mathbf{\Omega},t)\equiv1$)
and the detailed balance \begin{equation}
\hat{C}(t)\rho_{B}(\mathbf{J})=0,\label{DetBal}\end{equation}
which is responsible for bringing the system under study to the equilibrium
rotational Boltzmann distribution\begin{equation}
\rho_{B}(\mathbf{J})=(2\pi)^{-d/2}\exp\{-\mathbf{J}^{2}/2\}.\label{Boltz}\end{equation}
 Here $d=2$ for linear rotors and $d=3$ for spherical tops. Since
molecules are massive inertial particles, the relaxation operator
$\hat{C}(t)$ can depend upon $\hat{\mathbf{L}}$, $\mathbf{J}$,
and $\partial_{\mathbf{J}}$, but (in the absence of external fields)
is $\mathbf{\Omega}$-independent. This later requirement is a
direct consequence of isotropy of the rotational phase space.
It is important for the further consideration that $\hat{C}(t)$,
by its construction, is independent of the initial condition $\rho(\mathbf{J},\mathbf{\Omega},t=0)$.
To put it differently: once $\hat{C}(t)$ is chosen, it should describe
the time evolution of the probability density function $\rho(\mathbf{J},\mathbf{\Omega},t)$
for any initial condition. 

We shall further limit ourselves to the long-time (Markovian) evolution
of the probability density, i.e. assume that the timescale of interest
is much longer than the characteristic time of bath-induced fluctuations.
Thus, the master equation (\ref{kin1}) transforms into \begin{equation}
\partial_{t}\rho(\mathbf{J},\mathbf{\Omega},t)=\{-i\hat{\Lambda}-\hat{C}\}\rho(\mathbf{J},\mathbf{\Omega},t).\label{kin2}\end{equation}
 Here \begin{equation}
\hat{C}\equiv\int_{0}^{\infty}dt\hat{C}(t).\label{C}\end{equation}
 In addition, we concentrate on the evolution of the quantities
which depend on the angular momentum $\mathbf{J}$ but are independent
of the Euler angles $\mathbf{\Omega}$. Then, keeping in mind that
$\hat{C}$ is independent of $\mathbf{\Omega}$, we can integrate
Eq. (\ref{kin2}) over $\mathbf{\Omega}$ and arrive at the reduced
master equation

\begin{equation}
\partial_{t}\rho(\mathbf{J},t)=-\hat{C}\rho(\mathbf{J},t).\label{kinJ}\end{equation}

Here $\hat{C}$ can depend upon $\mathbf{J}$ and
$\partial_{\mathbf{J}}$.

Due to the fact that the relaxation operator (\ref{C}) obeys the
detailed balance (\ref{DetBal}), it is Hermitian. 
Let us denote its eigenfunctions and eigenvalues by $\rho_{B}(\mathbf{J})\Psi_{k}(\mathbf{J})$ and 
$\eta_{k}$, respectively ($k=0,1,2,...$). The eigenfunctions are assumed to be orthonormal, $\int d\mathbf{J}\rho_{B}(\mathbf{J})\Psi_{k}(\mathbf{J})\Psi_{m}(\mathbf{J})=\delta_{km}$. 
$\hat{C}$  possesses a
single eigenfunction $\rho_{B}(\mathbf{J})\Psi_{0}(\mathbf{J})$,  
$\Psi_{0}(\mathbf{J})\equiv1$, with the eigenvalue
$\eta_{0}=0$ (this is another way of saying that the Boltzmann distribution
is the right-hand eigenfunction of
$\hat{C}$ with zero eigenvalue) and all its other eigenfunctions
$\rho_{B}(\mathbf{J})\Psi_{k}(\mathbf{J})$ have positive eigenvalues $\eta_{k}$ ($k=1,2,...$).
If the eigenfunctions are known, we can evaluate any CF of interest.
Let us assume that, initially, the photofragments had a certain
non-equilibrium distribution, $\rho_{ne}(\mathbf{J})$. Then the angular
momentum CF $C_{J}(t)$, the averaged rotational energy $C_{S}(t)$,
and the rotational energy CF $C_{E}(t)$, are determined as follows:
\begin{equation}
C_{J}(t)=\frac{\left\langle \mathbf{JJ}(t)\right\rangle _{ne}}{\left\langle \mathbf{J}^{2}\right\rangle _{ne}}=\sum_{k=1}^{\infty}\frac{\left\langle \mathbf{J}\Psi_{k}(\mathbf{J})\right\rangle _{B}\left\langle \mathbf{J}\Psi_{k}(\mathbf{J})\right\rangle _{ne}}{\left\langle \mathbf{J}^{2}\right\rangle _{ne}}\exp\{-\eta_{k}t\}.\label{CJ}\end{equation}

\begin{equation}
C_{S}(t)=\frac{\left\langle E(t)\right\rangle _{ne}-\left\langle E\right\rangle _{B}}{\left\langle E\right\rangle _{ne}-\left\langle E\right\rangle _{B}}=\sum_{k=1}^{\infty}\frac{\left\langle E\Psi_{k}(\mathbf{J})\right\rangle _{B}\left\langle \Psi_{k}(\mathbf{J})\right\rangle _{ne}}{\left\langle E\right\rangle _{ne}-\left\langle E\right\rangle _{B}}\exp\{-\eta_{k}t\},\label{Sne}\end{equation}
\begin{equation}
C_{E}(t)=\frac{\left\langle EE(t)\right\rangle _{ne}-\left\langle E\right\rangle _{B}\left\langle E\right\rangle _{ne}}{\left\langle E^{2}\right\rangle _{ne}-\left\langle E\right\rangle _{B}\left\langle E\right\rangle _{ne}}=\sum_{k=1}^{\infty}\frac{\left\langle E\Psi_{k}(\mathbf{J})\right\rangle _{B}\left\langle E\Psi_{k}(\mathbf{J})\right\rangle _{ne}}{\left\langle E^{2}\right\rangle _{ne}-\left\langle E\right\rangle _{B}\left\langle E\right\rangle _{ne}}\exp\{-\eta_{k}t\}.\label{CBE}\end{equation}
We use the notation $\left\langle ...\right\rangle _{a}=\int d\mathbf{J}\rho_{a}(\mathbf{J})...$,
$a=B,\, ne$. 

The J-diffusion model \cite{gor66,rid69,McCl77}
and the standard rotational FPE \cite{hub72,McCo,mor82,McCl87}
predict the CFs $C_{S}(t)$ and $C_{E}(t)$ to be identical and single-exponential, even in the
case of arbitrary initial non-equilibrium distribution $\rho_{ne}(\mathbf{J})$.
The same does the Keilson-Storer model \cite{gel00,BurTe,kos06},
which contains the J-diffusion and the FPE models as a special case.
On the other hand, the simulations carried out in \cite{str06,str06a} 
demonstrate that (i) the averaged rotational energy $C_{S}(t)$ slows down significantly
with the increase of the rotational temperature of the photofragments  and 
(ii) $C_{S}(t)$ for hot non-equlibrium photofragments  differs dramatically from the equilibrium
rotational energy CF $C_{E}(t)$, manifesting thereby breakdown of the linear response description.

Eqs. (\ref{CJ})-(\ref{CBE}) offer a clear explanation of the failure of the standard
models of rotational relaxation to describe the above phenomena. Within
the J-diffusion model \cite{gor66,rid69,McCl77}, the (standard) rotational
FPE \cite{hub72,McCo,mor82,McCl87} and the Keilson-Storer model \cite{gel00,BurTe,kos06},
the angular momentum CF $C_{J}(t)$ is described by a single eigenfunction
$\Psi_{1}(\mathbf{J})$; $C_{S}(t)$ and $C_{E}(t)$ are also described
by a single, but different, eigenfunction, $\Psi_{2}(\mathbf{J})$. \cite{foot2}
Thus, irrespective of the initial condition $\rho_{ne}(\mathbf{J})$,
the rate of decay of these CFs is the same. 

If the eigenfunctions which
contribute significantly into the coefficients $\left\langle \Psi_{k}(\mathbf{J})\right\rangle _{ne}$,
$\left\langle E\Psi_{k}(\mathbf{J})\right\rangle _{ne}$, 
$\left\langle \Psi_{k}(\mathbf{J})\right\rangle _{B}$, and 
$\left\langle E\Psi_{k}(\mathbf{J})\right\rangle _{B}$
in Eqs. (\ref{CJ})-(\ref{CBE}) differ from each other,  
then the behavior of $C_{J}(t)$,  $C_{S}(t)$, and $C_{E}(t)$ 
under equilibrium and non-equilibrium conditions can be very different.
This is an indication that the standard rotational models must be generalized, 
allowing for the rate of rotational relaxation to be angular momentum dependent.
This requirement is easy to understand by using both classical and
quantum arguments. Indeed, let $\tau_{coll}$ be a characteristic
collision time. Such a collision can induce transition between the rotational
quantum states $j$ and $j+1$ with the frequency $\omega_{j+1,j}$
provided that \cite{bur74,BurTe} \begin{equation}
\omega_{j+1,j}\tau_{coll}=(\hbar j/I)\tau_{coll}\ll1.\label{mas}\end{equation}
The Massey parameter (\ref{mas}) tells us that the molecules possessing
relatively small angular momentum experience rotationally inelastic
collisions, while those possessing high enough angular momentum experience
$j$-conserving collisions. This means that the collision
rates must be $j$-dependent, and this is explicitly assumed in various
{}``fitting laws'', which are available in the literature for describing
$j$-resolved cross-sections. \cite{BurTe} If we are interested in CFs, which
are averaged over $j$, then, under equilibrium conditions, the majority
of rotational states with non-negligible Boltzmann factors are involved
in inelastic collisions. Clearly, if the rotationally
hot photofragments are produced via dissociation, then the contribution
due to $j$-conserving collisions will increase and must be properly
accounted for. If the classical picture is adopted, we can simply
state that the higher is the angular momentum, the more collisions
are necessary to randomize it. For example, the rate of the angular
momentum change in a binary collision collision is clearly $J$-dependent.
\cite{mul93} It is therefore not surprising that Gordon in his classical
paper \cite{gor66} on rotational relaxation suggested modifications
of his M-diffusion model toward $J$-dependent collision frequency.
This idea has further been elaborated in papers. \cite{rid69,McCl72,oui79,soo79,wyl80,gel98a}
However, the effects due to the $J$-dependent collision rates did
not receive much attention in the theory of rotational and orientational
relaxation, which deals primarily with ensemble-averaged rotational
and orientational CFs. Indeed, if we wish to describe these CFs under
equilibrium conditions, then (putting aside non-Markovian effects)
we can always introduce certain effective (averaged over $J$) collision frequencies or
relaxation rates. If we describe rotational relaxation in a wide range
of non-equilibrium initial conditions, these effective collision frequencies and 
rates are no longer applicable, and their explicit dependence on the
angular momentum must be taken into consideration.

On the basis of detailed  simulations, 
the authors of paper\cite{str06} have proposed the following physical interpretation 
of the slowing down of rotational relaxation of hot photofargments: highly rotationally energetic 
CN molecules push one (or several) argon atoms out of the solvation shell and, after that, rotate 
more or less freely before the restructuring of the shell occurs. This microscopic 
scenario corroborates entirely with our approach. This is  just another way of saying that highly 
rotationally excited molecules experience lower friction than their less energetic counterparts.

For the purposes of the description of CFs (\ref{CJ})-(\ref{CBE})
under non-equilibrium conditions we cannot, unfortunately,
use the M-diffusion model with $J$-dependent collision frequency
\cite{gor66} and related approaches \cite{rid69,McCl72,oui79,soo79,wyl80,gel98a},
since all these models assume $J$-conserving (adiabatic) relaxation mechanisms.
Neither can we straightforwardly generalize the J-diffusion or the
Keilson-Storer  model by introducing the $J$-dependent collision frequency, since
this procedure would violate normalization (and therefore conservation)
of the probability density (Eq. (\ref{norm})). The next Section is
aimed at developing the proper description.

\section{Generalized Fokker-Planck equation}

Without any loss of generality, the relaxation operator (\ref{C})
can be cast into the form of the generalized FPE operator:

\begin{equation}
\hat{C}=\partial_{\mathbf{J}}\hat{\xi}(\mathbf{J}+\partial_{\mathbf{J}}).\label{LiHo}\end{equation}
Here the generalized friction $\hat{\xi}$ is, in general,
an operator which depends upon $\mathbf{J}$ and $\partial_{\mathbf{J}}$.\cite{fre75,eva78,gel98} 
If the friction is constant ($\hat{\xi}=\xi=const$) then the standard
rotational FPE is recovered, which yields the single-exponential CFs \cite{hub72,McCo,mor82,McCl87,foot6}
\begin{equation}
C_{J}(t)=\exp\{-\xi t\},\,\,\, C_{S}(t)=C_{E}(t)=\exp\{-2\xi t\}.\label{all}\end{equation}

The concept of friction is fundamental for understanding the rotational  dynamics in solutions.\cite{coa96,mar96,mar00,bag99,str00} In the literature, a distinction has normally been made between the "dielectric" friction (which accounts for long-ranged interaction of polar solute molecules with a polar solvent) and "mechanical" friction (which is responsible for short-range anisotropic interactions). 
Since argon has been used as a solvent in the simulations carried out in \cite{str06}, we further focus on the "mechanical" friction. As has been explained in the previous Section, the drawback off all
standard models of rotational relaxation is the following: the relaxation rates are angular
momentum independent. To circumvent this problem, we employ the FPE (\ref{LiHo}) with the
angular momentum dependent friction (see, e.g., \cite{ris,sch00}
and references therein) and choose 

\begin{equation}
\hat{\xi}=\xi\frac{1}{1+b\mathbf{J}^{2N}}.\label{ksV}\end{equation}
Here the parameters $\xi,\, b\geq0$ are real, and $N$ is an integer.\cite{foot3} 
Such a functional form of the friction complies with qualitative
considerations presented in Sec. II. Furthermore, the same expression
has been suggested by Gordon \cite{gor66} as an extension of his
M-diffusion model and similar formulas are used for describing rotational
friction of "active" Brownian particles, which convert the energy of the environment
into the kinetic energy of their motion.\cite{sch00}    
It is necessary to emphasize that the parameters
$\xi$, $b$, and $N$, by their construction, must be independent
of the initial conditions. This means that if Eqs. (\ref{LiHo})
and (\ref{ksV}) describe rotational relaxation correctly, then they
must describe CFs in a wide range of initial conditions with fixed
$\xi$, $b$, and $N$.

For simplicity, we shall further restrict ourselves to the one-dimensional
case. That is, we assume that $\mathbf{J}$ is one-dimensional vector
(hereafter, the vector notation is therefore abandoned), and the equilibrium
Boltzmann distribution is given by Eq. (\ref{Boltz}) with $d=1$.
Furthermore, we assume that the non-equilibrium distribution can also
be written as a Boltzmann distribution, but at a different temperature
$T_{ne}$:\cite{foot4}
 \begin{equation}
\rho_{ne}(J)=(2\pi/\chi)^{-1/2}\exp\{-\chi J^{2}/2\},\,\,\,\,\chi\equiv T/T_{ne}.\label{ne}\end{equation}

We have used Maple 9.5 to numerically solve the FPE (\ref{LiHo})
with friction (\ref{ksV}) and calculate $C_{J}(t)$, $C_{S}(t)$,
and $C_{E}(t)$. The results of the calculations  are depicted in Fig. 1. As expected,
the decay times of all the CFs increase when we approach non-equilibrium
conditions (hot photofragments). All the non-equilibrium CFs $C_{J}(t)$, $C_{S}(t)$, and $C_{E}(t)$ differ considerably from each other as well as from their equilibrium counterparts (\ref{all}).
As is clearly seen, $C_{S}(t)$ calculated for hot photofragments 
deviates significantly from the equilibrium rotational energy CF $C_{E}(t)$ 
(compare the upper dashed line with the lower dotted line). This is in the agreement
with the results of papers, \cite{str06,str06a} which demonstrate inadequacy of the linear response theory in reproducing $C_{S}(t)$ far from equilibrium.

In order to get more insight into the behavior of CFs, it is helpful
to consider their integral correlation times 

\begin{equation}
\tau_{a}=\int_{0}^{\infty}dtC_{a}(t),\,\,\,\,\, a=J,\, E,\, S.\label{Tau}\end{equation}
 Now one more argument in favor of choosing algebraic friction (\ref{ksV})
is coming: as is shown in the Appendix, we can calculate Eqs. (\ref{Tau}) analytically.
The results read: 

\begin{equation}
\tau_{J}=\frac{1}{\xi}\left\{ 1+b\,\frac{(2N-1)!!}{\chi^{N}}\right\} ,\label{TauJ}\end{equation}
\begin{equation}
\tau_{S}=\frac{1}{2\xi}\left\{ 1+b\frac{(2N+1)!!}{N+1}\,\,\frac{1/\chi^{N+1}-1}{1/\chi-1}\right\} ,\label{TauS}\end{equation}
\begin{equation}
\tau_{E}=\frac{1}{2\xi}\left\{ 1+b\frac{(2N+1)!!}{N+1}\,\,\frac{(2N+3)/\chi^{N+1}-1}{3/\chi-1}\right\} .\label{TauE}\end{equation}
 If $b=0$, then we recover the standard FPE formulas. The presence
of $b\neq0$ causes the increase of the relaxation times. 

Eqs. (\ref{TauJ})-(\ref{TauE}) help us to reveal an important fact.
Let us assume that the $J$-dependence of friction (\ref{ksV}) is weak, that is  $b\ll1$.
If the initial conditions are close to
equilibrium ($\chi\sim1$), then the contribution due to the 
$J$-dependent friction in Eqs. (\ref{TauJ})-(\ref{TauE}) is
proportional to $b$ and is therefore  small. This observation
supports the use of the models with constant relaxation rates (the
J-diffusion model, \cite{gor66,rid69,McCl77} the standard rotational
FPE, \cite{hub72,McCo,mor82,McCl87} and the Keilson-Storer model \cite{gel00,BurTe,kos06})
for the description of molecular reorientation under equilibrium conditions.
Suppose that the initial conditions are highly non-equilibrium, so
that the characteristic temperature of the photofragments is much
higher than the bath temperature, $\chi=T/T_{ne}\ll1$. Then the contribution
of the second term in Eqs. (\ref{TauJ})-(\ref{TauE}) is determined
by the factor of $b/\chi^{N}$, which is no longer small even if
$b\ll1$.\cite{foot5}  Thus 
\begin{equation}
T_{ne}\sim T/\sqrt[N]{b} \label{thresh}\end{equation}
delivers the threshold value of the non-equilibrium temperature (or rotational energy), for which the slowing down of rotational relaxation becomes significant. The existence of such a threshold temperature is clearly demonstrated in Ref.\cite{str06a} The farther we are from equilibrium (the higher is $\chi$), the stronger is
the effect due to the $J$-dependence of friction. This
finding corroborates a general paradigm of statistical physics and
non-equilibrium thermodynamics, which emphasizes the growing role of fluctuations far
from equilibrium. 

The above considerations explain why the linear-response theory, which identifies the 
non-equilibrium averaged rotational energy $C_{S}(t)$ with equilibrium rotational energy fluctuations $C_{E}(t)$, breaks down far from equilibrium. If we are close to equilibrium, then the contribution of the high-$J$ states into rotational relaxation is minor. Thus the equilibrium fluctuations sample adequately the angular momentum exchange between the solute and solvent molecules. If we move far from equilibrium, then rotational relaxation spreads over the high-$J$ states, and equilibrium fluctuations are unable to adequately sample the angular momentum exchange.     

Once Eqs. (\ref{TauJ})-(\ref{TauE}) are available, we can attempt
to make more quantitative comparison of the present theory with the
results of simulations reported in. \cite{str06} If our approach
grasps essential physics of the phenomenon, we shall be able to fit
all the simulated  $C_{S}(t)$ by a single set of the parameters $N$, $\xi$,
and $b$. We have thus proceeded as follows. We have numerically
calculated the integral relaxation times of $C_{S}(t)$ simulated
at different $T_{ne}$. By combining the obtained dimensionless values of $\tau_{S}$
at equilibrium ($T_{ne}=T=120$K) and at a certain temperature $T_{ne}>T$
we have used Eq. (\ref{TauS}) to calculate the corresponding parameters
$\xi$ and $b$ for $N=1,\,2$. The results of these calculations are
summed up in Table 1. If our approach is self-consistent, the values
of $\xi$ and $b$ must be more or less the same for any $T_{ne}$.
As is seen from the Table, this is certainly not the case for $N=1$,
while the calculations with $N=2$ exhibit a remarkable consistency.
Thus we can speculate that the FPE (\ref{LiHo})
with friction (\ref{ksV}) at $N=2$ describes the simulated $C_{S}(t)$ rather
adequately. The so calculated $C_{S}(t)$, along with their simulated counterparts
\cite{str06} are depicted in Fig. 2. The overall agreement is satisfactory,
but not perfect. There exists a number of reasons for such a deviation. First,
the simulated CFs clearly exhibit short-time non-Markovian behavior.
In order to describe this, the present FPE should be generalized to
account for memory effects (see, e.g., \cite{gel96a}). Such a generalization is
necessary, for example, for reproducing the short-time ($\sim100$fs) coincidence 
of the simulated $C_{S}(t)$ at different $T_{ne}$.\cite{str06} Second, in
order to calculate the integral relaxation times (\ref{TauJ})-(\ref{TauE})
analytically, we have switched to the one-dimensional model, while the
actual description for linear photofragments must be two-dimensional.
Third, as is well known, the FPE itself may not be very good for reproducing
rotational relaxation in liquids \cite{LB84}, so that more refine approaches
might be necessary. The present paper undertakes the first step toward
understanding non-equilibrium  rotational relaxation. The message is as follows: the friction, or the collision
frequency, or the relaxation rate must be taken $J$-dependent in
order to adequately describe rotational relaxation under highly non-equilibrium
conditions.

\section{Conclusion}

Recently, the groups of Bradforth and Stratt have published the results
of the combined theoretical and experimental study of rotational relaxation
of CN-photofragments in a condensed phase. \cite{str06,str06a} They
show via molecular dynamics simulations that the time evolution of
the averaged rotational energy, $C_{S}(t)$, (i) slows 
down dramatically with the increase of the rotational temperature of the incipient
photofragments and (ii) deviates significantly from the equilibrium
rotational energy CF, $C_{E}(t)$, manifesting the linear response breakdown.
To comprehend and explain this
unusual behavior, we have developed a theory of rotational relaxation
under non-equilibrium conditions, by extending the standard FPE approach
to account for angular momentum dependent friction. We have calculated the
angular momentum CF $C_{J}(t)$, as well as $C_{S}(t)$ and $C_{E}(t)$,
and compared them with their simulated counterparts. \cite{str06}
We have shown that $J$-dependence of rotational friction is responsible
for the observed slowing down of rotational relaxation and failure of the linear response description, provided the effective non-equilibrium rotational temperature exceeds the threshold value (\ref{thresh}). The farther the system is from the equilibrium, the stronger is the effect.

\begin{acknowledgments}
We are grateful to Guohua Tao and Richard M. Stratt for providing us with the numerical data on the calculated CFs and measured anisotropies which have been published in\cite{str06}, for sending us a manuscript of paper\cite{str06a} prior to publication, and for helpful discussions.     This work was partially supported by the American Chemical Society Petroleum Research Fund (44481-G6)
and  General Research Board summer award from the University of Maryland.

\end{acknowledgments}

\appendix
\section{Calculation of the integral relaxation times }

We start from the one-dimensional FPE with the angular momentum dependent
friction $\xi(J)$,\begin{equation}
\partial_{t}\rho(J,t)=\partial_{J}\xi(J)(J+\partial_{J})\rho(J,t).\label{fpe1}\end{equation}
Evidently, a CF of the functions $A(J)$ and $B(J)$ can be evaluated
via Eq. (\ref{fpe1}) as follows: \begin{equation}
C_{AB}(t)=\left\langle A(J)B(J(t))\right\rangle _{ne}-\left\langle A(J))\right\rangle _{ne}\left\langle B(J)\right\rangle _{B}=\int_{-\infty}^{\infty}dJB(J)\rho(J,t)\label{Cab}\end{equation}
(the notation is identical to that used in Eqs. (\ref{CJ})-(\ref{CBE})).
The initial condition to Eq. (\ref{fpe1}) reads then:\[
\rho(J,0)=A(J)\rho_{ne}(J)-\left\langle A(J)\right\rangle _{ne}\rho_{B}(J).\]
Here the initial non-equilibrium distribution is determined by Eq.
(\ref{ne}), and the Boltzmann equilibrium distribution is defined
via Eq. (\ref{Boltz}) with $d=1$. Let \[
\tau_{AB}=\int_{0}^{\infty}dt\left\{ \left\langle A(J)B(J(t))\right\rangle _{ne}-\left\langle A(J))\right\rangle _{ne}\left\langle B(J)\right\rangle _{B}\right\} \]
be the integral relaxation time for the above CF. Then, integrating
Eq. (\ref{fpe1}) over time and introducing the quantity \[
\eta(J)=\int_{0}^{\infty}dt\rho(J,t),\]
we can express $\tau_{AB}$ through $\eta(J)$ as follows:\begin{equation}
\tau_{AB}=\int_{-\infty}^{\infty}dJB(J)\eta(J).\label{TauAB}\end{equation}
 Here $\eta(J)$ obeys the differential equation \begin{equation}
-\left\{ A(J)\rho_{ne}(J)-\left\langle A(J)\right\rangle _{ne}\rho_{B}(J)\right\} =\partial_{J}\xi(J)(J+\partial_{J})\eta(J).\label{fpe2}\end{equation}
Eq. (\ref{fpe2}) must be solved with the boundary conditions $\eta(\pm\infty)=0$.
It is convenient to introduce the quantity $\widetilde{\eta}(J)$
via the expression\begin{equation}
\eta(J)\equiv\widetilde{\eta}(J)\rho_{B}(J).\label{til}\end{equation}
Upon the insertion of the above equation into Eq. (\ref{fpe2}), we
get: \begin{equation}
-\frac{1}{\rho_{B}(J)\xi(J)}\int_{0}^{J}dJ'\left\{ A(J')\rho_{ne}(J')-\left\langle A(J)\right\rangle _{ne}\rho_{B}(J')\right\} =\partial_{J}\widetilde{\eta}(J)\label{fpe3}\end{equation}
(the integration limits have been chosen to comply with the boundary
conditions $\eta(\pm\infty)=0$). As is seen, Eq. (\ref{fpe3}) can
be integrated, in principle, for any $\rho_{ne}(J)$, $\xi(J)$, and
$A(J)$. To simplify the subsequent presentation, we shall limit ourselves
to the case when\begin{equation}
B(J)=J^{m},m=1\,\textrm{or}\,2.\label{AB}\end{equation}
Inserting this expression into Eq. (\ref{TauAB}), using the definition
(\ref{til}) and integrating by parts yields then\begin{equation}
\tau_{AB}=\int_{-\infty}^{\infty}dJJ^{m-1}\rho_{B}(J)\partial_{J}\widetilde{\eta}(J),\,\,\, m=1\,\textrm{or}\,2.\label{TauAB1}\end{equation}
Combining Eqs. (\ref{fpe3}) and (\ref{TauAB1}), we get:\begin{equation}
\tau_{AB}=-\int_{-\infty}^{\infty}dJJ^{m-1}\frac{1}{\xi(J)}\int_{0}^{J}dJ'\left\{ A(J')\rho_{ne}(J')-\left\langle A(J)\right\rangle _{ne}\rho_{B}(J')\right\} ,\,\,\, m=1\,\textrm{or}\,2.\label{fpe4}\end{equation}
Using the explicit form (\ref{ksV}) of the angular momentum friction
and integrating Eq. (\ref{fpe4}) by parts gives, finally:

\begin{equation}
\tau_{AB}=\frac{1}{\xi}\int_{-\infty}^{\infty}dJ\left(\frac{J^{m}}{m}+b\frac{J^{m+2N}}{m+2N}\right)\left\{ A(J)\rho_{ne}(J)-\left\langle A(J)\right\rangle _{ne}\rho_{B}(J)\right\} ,\,\,\, m=1\,\textrm{or}\,2.\label{fpe5}\end{equation}
This is the formula which has been used for the calculation of integral
relaxation times (\ref{TauJ})-(\ref{TauE}).

\clearpage
\begin{table}[b]
\caption{ The dimensionless parameters $\xi$ and $b$ calculated via
Eq. (\ref{TauS}) from the integral relaxation times $\tau_{S}$ (given
in units of $\tau_{r}$) simulated at different temperatures $T_{ne}$
(in K). Integral relaxation time $\tau_{S}=1.77$
at equilibrium temperature $120$K.} 
\label{tbl:table1}
\begin{tabular}{ccc|ccc|cc}
\hline\hline
\multicolumn{2}{c}{Simulation\cite{str06} } &&  \multicolumn{2}{c}{$N=1$} && \multicolumn{2}{c}{$N=2$}   \\
\hline
$T_{ne}$&$\tau_{S}$ && $\xi$& $b$&&  $\xi$& $b$ \\
\hline
$1917$&$5.29$    &&  $0.39$&$0.12$  && $0.29$& $0.0015$ \\
$2395$& $7.73$   &&  $0.44$& $0.18$ && $0.29$& $0.0017$ \\
$2875$& $11.25$ &&  $0.53$& $0.29$ && $0.29$& $0.0018$
\\
\hline
\hline
\end{tabular}
\end{table}

\clearpage

\begin{figure}
\includegraphics[keepaspectratio,totalheight=10cm]{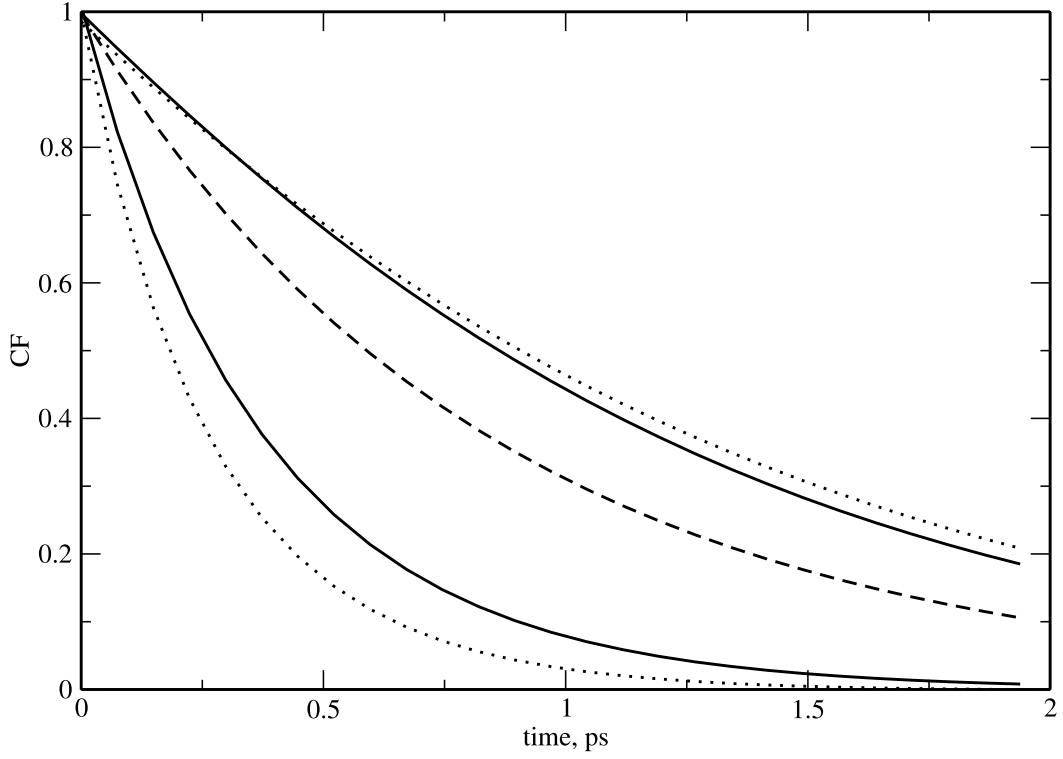}
\caption{ 
Angular momentum CF $C_{J}(t)$ (full lines), averaged energy
$C_{S}(t)$ (dashed lines) and rotational energy CF $C_{E}(t)$ (dotted
lines) for algebraic friction (\ref{ksV}) with $N=1$, $\xi=1$,
and $b=0.29$. The lower CFs are calculated for equilibrium conditions
($\chi=T/T_{ne}=1$) and the upper CFs are computed for hot photofragments
($\chi=0.1$). 
 }
\end{figure}

\clearpage
\begin{figure}
\includegraphics[keepaspectratio,totalheight=12cm]{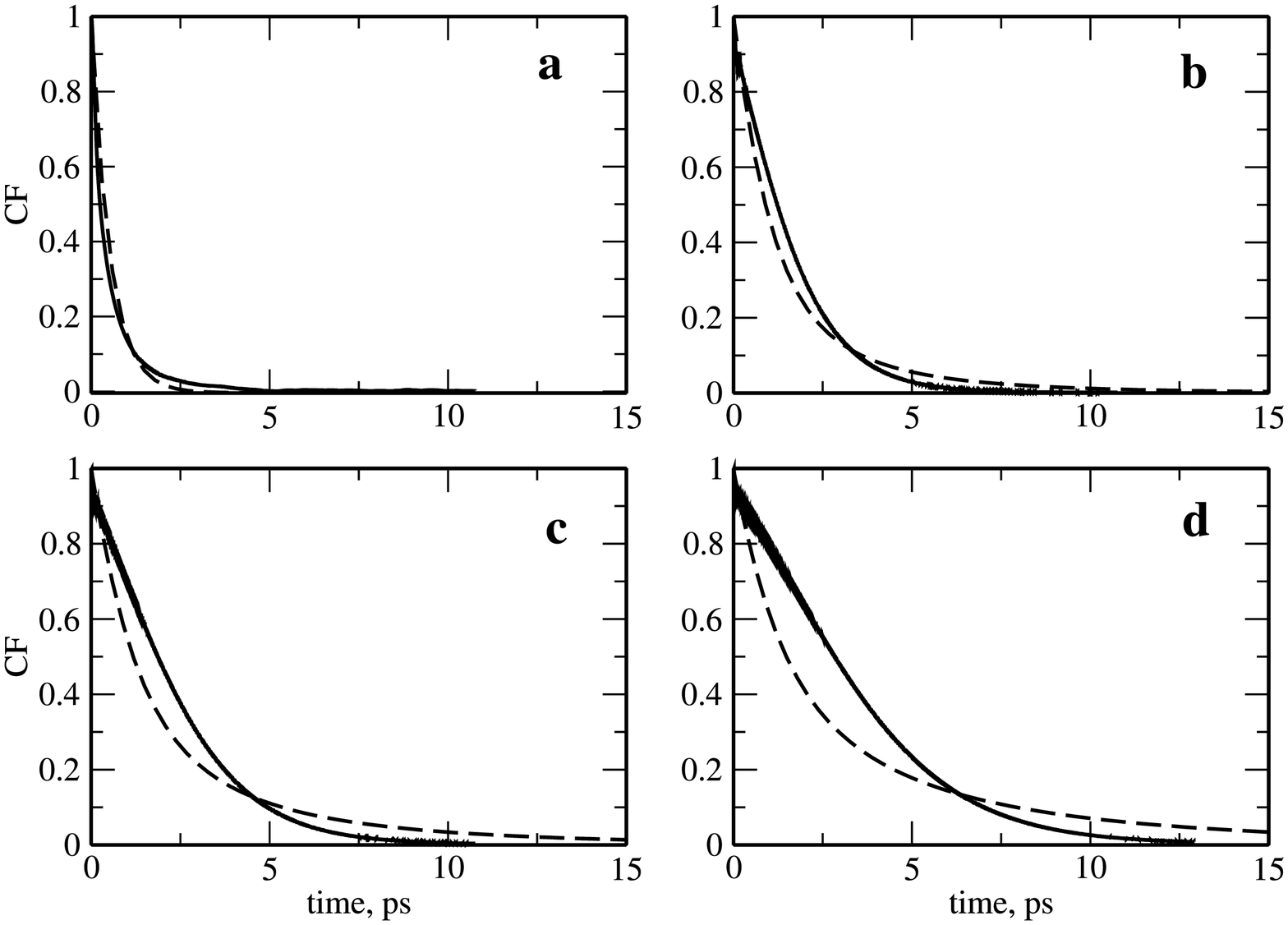}
\caption{ 
Averaged energy CF $C_{S}(t)$ for $T_{ne}=120$K (a), $1917$K
(b), $2395$K (c) and $2875$K (d). Full lines correspond to the molecular
dynamics simulations, \cite{str06} and dashed lines display the results
of the FPE calculations with algebraic friction (\ref{ksV}).  $N=2$ and
the parameters $\xi$ and $b$ are listed in Table 1. 
}
\end{figure}


\begin{thebibliography}{20}
\bibitem{foot1}This assumption, in principle, can be relaxed but it is adequate.
Since photodissociation is normally quite fast (hundreds of femtoseconds\cite{zew94,zew01})
on the rotational dynamics timescale,
the ensemble of photofragments can be thought of as being instantaneously
injected into the equilibrium heat bath.
\bibitem{foot6} Strictly speaking, the averaged rotational energy CF, 
$
\left\langle E(t)\right\rangle _{ne}-\left\langle E\right\rangle _{B}
$, 
is identically zero under equilibrium conditions ($ne\rightarrow B$). However, the normalized CF $C_{S}(t)$, as defined by Eq. (\ref{Sne}), yields $\exp\{-2\xi t\}$ in the limit $ne\rightarrow B$.
\bibitem{foot3} There is, perhaps, no good reason to consider high values of $N$
in Eq. (\ref{ksV}), so that the first few $N=1,2,3$ are meaningful.
\bibitem{foot4} A more physically justified parametrization of  $\rho_{ne}(\mathbf{J})$ can be found in Refs. \cite{zew94,gel00,gel01,gel02}
\bibitem{foot5} Note that Eqs. (\ref{TauJ})-(\ref{TauE}) predict that rotational relaxation in non-equilibrium cold ensembles ($\chi=T/T_{ne}>1$) is faster than in equilibrium ensembles ($\chi=T/T_{ne}=1$). Since the parameter $b$ is assumed to be small, this effect is not so pronounced, in general.
\end{thebibliography}
\end{document}